# $1/f^2$ spectra of decoherence noise on $^{75}$As nuclear spins in bulk GaAs

*Susumu Sasaki[1,2], #Takanori Miura[3], #Kohsuke Ikeda[3], #Masahiro Sakai[3],

Takuya Sekikawa[3], Masaki Saito[3], Tatsuro Yuge[4], Yoshiro Hirayama[5]

[1] *Materials Science Program, Niigata University, Niigata 950-2181, Japan*
[2] *Japan Agency for Medical Research Development, Tokyo 100-0004, Japan*
[3] *Graduate School of Science and Technology, Niigata University, Niigata 950-2181, Japan*
[4] *Department of Physics, Shizuoka University, Shizuoka 422-8529, Japan*
[5] *Department of Physics, Tohoku University, Sendai 980-8578, Japan*
#These authors contributed equally to this work.

To identify the decoherence origin, frequency spectra using multiple π-pulses have been extensively studied. However, little has been discussed on how to define the spectral intensities from multiple-echo decays and how to incorporate the Hahn-echo $T_2$ in the noise spectra. Here, we show that experiments based on two theories solve these issues. With the previous theory clarifying that the spectral intensity should be given as the decay in the long-time limit, the intensity can be deduced without experimental artifacts usually entailed in the initial process. The other is the fluctuation-dissipation theory, with which the Hahn-echo $T_2$ is utilized as the zero-frequency limit of the noise spectrum and as an answer to the divergent issue on the $1/f^n$ noises. As a result, arsenic nuclear spins are found to exhibit $1/f^2$ dependences over two orders of magnitude in all the substrates of un-doped, Cr-doped semi-insulating and Si-doped metallic GaAs at 297 K. The $1/f^2$ dependence indicates single noise source that is characterized by the characteristic frequency $f_c^{un}$=170±10 Hz, $f_c^{Cr}$=210±10 Hz and $f_c^{Si}$ =460±30 Hz. These $f_c$ values are explained by a model that the decoherence is caused by the fluctuations of next-nearest-neighboring nuclear spins.

 To realize quantum computing, electron-, nuclear-, and pseudo-spins have been extensively studied as promising candidates of quantum bits. One of the greatest findings is to apply multiple π-pulses for not just enhancing the coherence time but presenting decoherence-noise spectra. In these studies, some report the usual $1/f$ and others $1/f^3$ spectra[1], where very little has been discussed on how to deduce the spectral intensities from their raw data: the multiple-echo decay. Moreover, the Hahn-echo $T_2$, on which the coherence enhancement is based, has not been incorporated in noise spectra.

 In this paper, we demonstrate that experiments based on two theories solve these issues. In our previous theory, it is proved that the spectral intensity should be defined as the multiple-echo decay in the long-time limit[2]. We take advantage of this definition, which is free from experimental artifacts usually entailed in the initial decays. Experimentally, we employ a multiple alternating-phase π-pulse sequence instead of the usually employed Carr-Purcell-Meiboom-Gill (CPMG)[1], resulting in multiple-echo decays for shorter π-pulse intervals. We also take into account the fluctuation-dissipation theorem[3] that predicts that the Hahn-echo $T_2$ reflects the zero-frequency intensity. This indicates that the Hahn-echo $T_2$ generally solves the divergent issue on the $1/f^n$ noises and validates the spectrum as the maximum value at zero frequency. As a result, $^{75}$As nuclear spins are found to exhibit, for all the substrate of un-doped, Cr-doped semi-insulating and Si-doped metallic GaAs, not the usual $1/f$ noise but a Loerntzian $1/f^2$ spectrum at 297 K. The $1/f^2$ dependences are characterized by the frequency $f_c{}^{un}$=170±10 Hz and $f_c{}^{Cr}$=210±10 Hz. These $f_c$ values are explained by a model that the decoherence is caused by the fluctuations of next-nearest-neighboring nuclear spins. For the metallic GaAs, the somewhat larger value $f_c{}^{Si}$=460±30 Hz can be understood that the spin fluctuation of doped carriers enhances the dipolar coupling between the nearby nuclear spins, and hence the increase in $f_c$.

**Theories**

Before discussing the results, we briefly outline the theories[2,3] on which the present experiments are based. The fluctuation-dissipation theorem proves that the spin-lattice relaxation time $T_1$ and the Hahn-echo $T_2$ are given as[3]

$$\frac{1}{T_1} = \frac{\gamma_n^2}{2} J_\perp(\omega) \tag{1}$$

$$\frac{1}{T_2} = \frac{\gamma_n^2}{2} J_\parallel(0) \tag{2}$$

where $\gamma_n$ is the gyromagnetic ratio of the nuclear spin, $J_\perp(\omega)$ [$J_\parallel(0)$] is the spectrum intensity due to the fluctuating magnetic field $\delta B_{x,y}(t)$ [$\delta B_z(t)$] acting on the nuclear spin where a static magnetic field $B_z$ is applied in the $z$-direction. The $\omega$ in Eq. (1) is the resonance frequency given as $\gamma_n B_z$ of the order of 10-100 MHz, whereas $1/T_2$ is given by the *zero*-frequency $J_\parallel(0)$. The inclusion of the Hahn-echo $T_2$ solves the "divergent" issue[4] of the $1/f^n$ noise in the zero-frequency limit.

To extend Eq.(2) into *non-zero* frequency, we consider a decay of multiple echoes of spin (nuclear- or electron- or pseudo-spin) as a result of applying a sequence of multiple π-pulses. In the long-time limit, it is clarified[2] that the multiple-echo decay intensity $s_{APCP}(t)$ exhibits an exponential function as

$$s_{APCP}(t) \to \exp(-\frac{t}{T_2^L}) \tag{3}$$

This relation is found to hold for any multiple π-pulse sequence so long as all the π-pulse intervals (2τ) are equal and for the cases that $T_1 \gg T_2$. With the $T_2^L$ defined as the decay in the long-time limit [Eq.(3)], Eq. (2) is extended into *non-zero* frequencies as

$$\frac{1}{T_2^L(f)} = \frac{\gamma_n^2}{2} \frac{8}{\pi^2} \sum_{m=0}^{\infty} \frac{1}{(2m+1)^2} J_\parallel((2m+1)f) \tag{4}$$

where the frequency $f$ is given as $f = 1/4\tau$. The factor $1/(2m+1)^2$ in the $J_\parallel((2m+1)/4\tau)$ term allows us to approximate Eq.(4) by the fundamental ($m$=0) harmonic $J_\parallel(1/4\tau)$ to a good approximation as

$$\frac{1}{T_2^L} \approx \frac{\gamma_n^2}{2} \frac{8}{\pi^2} J_\parallel(f) \tag{5}$$

From the mathematical formula $\sum 1/(2m+1)^2 = \pi^2/8$, the zero-frequency limit of Eq.(4) is reduced to Eq.(2). Alternatively, the Hahn-echo $T_2$ that reflects $J_\parallel(0)$ [Eq.(2)] can be extended into non-zero frequency [Eq.(4)] as the generalized coherence time $T_2^L$.

**Experiments**

All the measurements are done for commercially available substrate of un-doped, Cr-doped semi-insulating and Si-doped metallic GaAs (P/N N-5181-001, VAT, Switzerland) at 297 K using a standard NMR spectrometer which we built up and customized to ourselves. The static magnetic field $B_z$ of 6.166 T generated by a superconducting magnet (Oxford 300/89) is applied perpendicular to [0 0 1] surface of the substrate. Frequency spectra of $^{75}$As-NMR show a single line without any quadrupole splitting[SI-1], with the full width at half maximum of 5.8 kHz and 6.4 kHz, and 6.8 kHz for un-doped, Cr-doped semi-insulating and Si-doped metallic GaAs substrate, respectively. To obtain multiple spin-echo decays or the Hahn-echo $T_2$, we irradiate sufficiently strong power roughly 300 W or more, resulting in the π/2-pulse width of 2.2 μs for un-doped GaAs, and 8.0 μs for both Cr-doped semi-insulating and Si-doped metallic GaAs. From these pulse widths, the frequency range that the irradiated pulses can manipulate is estimated to be 114 kHz (> 5.8 kHz) for un-doped GaAs and 31.3 kHz (> 6.4 or 6.8 kHz) for both Cr-doped semi-insulating and Si-doped metallic GaAs. This proves that the irradiated pulses manipulate all the nuclear spins observed in the frequency spectra. The assumption on which Eqs. (2-5) are based is $T_1 \gg T_2$,[2] which is experimentally proved to hold in the present cases[SI-5].

Instead of the widely-used CPMG pulse sequence, we employ alternating-phase Carr-Purcell (APCP) pulses that alternate the polarity of multiple echoes (Fig.1a), enabling us to obtain $T_2^L$ for shorter $2\tau$'s. Also employed are the phase-cycling technique to eliminate ring-down noises from spin-echo signals, and the quadrature detection to increase signal-to-noise ratio[5]. For a multi-exponential decay where a single-exponential behavior is theoretically expected, the long-time limit behavior is generally considered intrinsic since the initial decay depends on experimental artifacts. In the present cases, the slopes of initial decays are found to depend on the homogeneity of alternating magnetic field while $1/T_2^L$ is intact (Fig.1c).

**Results**

Figure 2 shows multiple-echo decays for various $2\tau$'s for $^{75}$As nuclear spin in un-doped [Fig.2 (a)], Cr-doped semi-insulating [Fig.2 (b)] and Si-doped GaAs [Fig.2 (c)] substrate. The shorter the $2\tau$, the longer the $s_{APCP}(t)$ persists, which is well known as the dynamical decoupling[1]. To deduce $T_2^L$ in the long-time limit, triple-exponential functions are found[SI-3, SI-4] to consistently reproduce all the decays for various $2\tau$'s. To be consistent with $T_2^L$ at nonzero frequencies, we define the Hahn-echo $T_2$ not as the usually employed initial decay but as the exponential function in the long-time limit[2, SI-2].

Figure 3 shows the noise spectra $J_\parallel(f)$ represented by $1/T_2^L$ as a function of frequency $f = 1/4\tau$ for un-doped [Fig.3 (a)], Cr-doped semi-insulating [Fig.3 (b)] and Si-doped GaAs [Fig.3 (c)] substrate. To reproduce the data both at zero- and high-frequency, the fitting function should be described as

$$\frac{1}{T_2^L(f)} = \frac{A}{1+(f/f_c)^n} + B \quad (6)$$

where $A$ is given by the long-time limit of the Hahn-echo decay as $A^{un}$=1.60±0.05 kHz for un-doped GaAs, $A^{Cr}$=1.70±0.08 kHz for Cr-doped semi-insulting GaAs and $A^{Si}$=0.85±0.04 kHz for Si-doped metallic GaAs. $1/T_2^L$ values at high frequencies give the $B$ value as $B^{un}$=7.3±0.4 Hz, $B^{Cr}$=24.7±0.9 Hz and $B^{Si}$=15.5±0.7 Hz. For further investigations, we subtract the $B$ term from all the $1/T_2^L$ values. As the broken curves in Fig. 3 show, the $f$-dependence in Eq. (6) are found to be $n$=2.00±0.01 for all the substrate which holds over two orders of magnitude. As a result, the noise spectra are reproduced by Eq. (6) with the fitting parameter $f_c^{un}$=170±10 Hz, $f_c^{Cr}$=210±10 Hz and $f_c^{Si}$=460±30 Hz.

**Discussions**

First, we discuss the noises dominant at lower frequencies. From the basics of Fourier transformation, the first term in Eq.(6) indicates that the noise is due to a single source[4] with the characteristic frequency $f_c$ as in $<\delta B_z(t+t')\delta B_z(t)> \propto \exp(-f_c t')$. To be noted is the $f_c^{un}$ and $f_c^{Cr}$ values of ~200 Hz that are extremely smaller than the fluctuation frequencies of other noise spectra[1]. For un-doped and semi-insulating GaAs, neither spins of doped carriers nor magnetic impurities are likely. Here we show that $\delta B_z(t)$ is caused by the fluctuation of surrounding nuclear spins from a rough estimation on the $f_c^{un}$ and $f_c^{Cr}$ values. Dipolar field on $^{75}$As due to the nearest neighbor $^{69}$Ga nuclear spins is estimated that $<^{69}B_{dip}> = ^{69}\mu/r^3 = \pm 0.68$ Gauss, where $r$ is the distance between $^{69}$Ga and $^{75}$As nuclear spins. Similarly, the dipolar field by the nearest neighbor $^{71}$Ga is given as $<^{71}B_{dip}> = \pm^{71}\mu/r^3 = \pm 0.87$ Gauss. The dipolar field due to the nearest-neighbor Ga nuclear spins result in the de-phasing of $^{75}$As nuclear spin with the frequency of $^{75}\gamma<^{69}B_{dip}> \sim 1.00$ kHz and $^{75}\gamma<^{71}B_{dip}> \sim 1.27$ kHz. Apparently, the estimated values are significantly greater than the $f_c^{un}$ value of 170±10 Hz. If we suppose that the de-phasing is caused by the next-nearest neighbor dipolar field that $<^{75}B_{dip}> = \pm^{75}\mu/r^3 = \pm 0.11$ Gauss, we obtain the frequency that $^{75}\gamma<^{75}B_{dip}> \sim 163$ Hz, which is very close to the experimental result that $f_c^{un}$=170±10 Hz. In an analogy with the RKKY interaction[6], the doped-carrier spins are likely to enhance the dipolar coupling between the adjacent Ga and As nuclear spins, resulting in the de-phasing frequency of the $^{75}$As nuclear spins. This is consistent with the fact that $f_c^{Si}$ =460±30 Hz is greater than $f_c^{un}$=170±10 Hz and $f_c^{Cr}$=210±10 Hz. Thus, the decoherence noise in the lower frequencies are likely to come from the next-nearest-neighboring nuclear spins.

Next we consider the origin of the higher frequency $B$ terms. In the limit that $f_c \to \infty$, the Lorentzian spectrum generally exhibits a frequency-independent function, the intensity of which approaches zero. Thus, the higher-frequency $B$ term can be viewed as another single noise of the Lorentzian spectrum with a characteristic frequency much larger than $f_c^{un}$, $f_c^{Cr}$ and $f_c^{Si}$. As a candidate for the higher frequency $B$ terms, we consider the possibility of the Fermi-contact interaction[7] where the spins of carriers contribute to the $^{75}$As nuclear-spin decoherence. For un-doped or semi-insulating [Si-doped] GaAs, the average distance between adjacent carriers is of the order of $10^{-5}$ [$10^{-6}$] cm from the carrier concentration of $10^{15}$ [$10^{18}$] cm$^{-3}$. Since the free-electron model gives us the Fermi velocity of roughly $10^3$ [$10^4$] m/s for un-doped or semi-insulating [Si-doped] GaAs, the frequency with which carriers are diffracted by $^{75}$As is estimated to be of the order of $10^5$ [$10^7$] kHz, which is sufficiently larger than the low-frequency $f_c^{un}$ or $f_c^{Cr}$ [$f_c^{Si}$]. This indicates that the higher frequency $B$ term is contributed by the Fermi-contact interaction but dominated by other mechanisms such as inhomogeniety of doped atoms.

Finally, we show how the noise spectra are dependent on the definition of the spectral intensity and on the inclusion of the Hahn-echo $T_2$. Figure 4 shows noise spectra without the Hahn-echo $T_2$. In this case, the spectra are fit by $A/(f/f_c)^n + B$. If we take the initial slope $T_2^S$ as the spectral intensity, the spectra exhibit the usual $1/f^{1\pm 0.3}$ spectra. On the other hand, if we take the $T_2^L$ as the spectral intensity, all the spectra approximately result in $1/f^{2\pm 0.3}$ dependence, even in the absence of the Hahn-echo $T_2$. This proves the validity of $T_2^L$ in the definition of the spectral intensity. In Fig. 5, we show the noise spectra based on $T_2^S$ with the Hahn-echo $T_2$ incorporated. In contrast to Fig. 4(a), all the spectra are exhibit $1/f^{2\pm 0.3}$ dependence, which proves the importance of the Hahn-echo $T_2$.

**Conclusions**

To obtain the decoherence noise spectrum from a multiple spin-echo decay, we have followed the previous paper that the intensity should be defined as $1/T_2^L$, the slope of the long-time limit. This method is found to avert the experimental artifacts usually entailed in the spin-echo decays. Based on the fluctuation-dissipation theorem, we have utilized the Hahn-echo $T_2$ to validate the noise spectrum as the zero-frequency intensity, and to solve the apparent divergent behavior of the $1/f^n$ noises. With the help of these two theories, we have found that $^{75}$As nuclear spins exhibit $1/f^2$ dependences over two orders of magnitude in all the substrates of un-doped, Cr-doped semi-insulating and Si-doped metallic GaAs at 297 K. The $1/f^2$ dependence indicates single noise source that is characterized by the frequency that $f_c^{un}$=170±10 Hz, $f_c^{Cr}$=210±10 Hz and $f_c^{Si}$ =460±30 Hz.

We have shown from a rough estimation that the decoherence is caused by the fluctuations of next-nearest-neighboring nuclear spins. We have also proved the validity of $T_2^L$ as the spectral intensity and the inclusion of the Hahn-echo $T_2$.

**Acknowledgments**

   The authors acknowledge financial supports from Tohoku University CSRN and KAKENHI Grants (Nos. 26287059 and 15H05867). S.S. and Y.H. are grateful to stimulated discussions in the meetings of Cooperative Research Project of Research Institute of Electrical Communication, Tohoku University. S.S. is also supported partly by the joint research program of Research Institute of Electrical Communication (RIEC) Tohoku University, the collaboration program of Institute of Materials and Systems for Sustainability (IMaSS), Nagoya University, KAKENHI grants (No 19H02580), the Naito Scholarship Foundation and the Yamgaguchi Educational and Scholarship Foundation.

**Author contributions**

   SS presented ideas and supervised all experiments and analyses, and wrote the paper. With SS's instructions, TM, KI, MS carried out experimental and analytical works with equal contributions. Sekikawa and Saito checked all the estimated values. TY performed theoretical calculations and proved SS's prediction on the long-time behavior. All authors, particularly YH, took part in discussing results and analyses in detail, and in editing the manuscript.

**Competing financial interests**

 The authors declare no competing financial interests.

**Materials and correspondence**
*Susumu Sasaki (susumu @ niigata-u.ac.jp)

**Figure 1| APCP pulse sequence, multiple echoes and robustness of long-time behaviors.** Top: Schematic illustration that an APCP sequence (blue) with the π-pulse interval of 2τ causes a train of multiple spin-echoes (green). Subscripts *x* or *-x* indicates the axis in the rotating frame along which the RF pulses are applied. Middle: Multiple spin-echo signals (red) under an APCP sequence with 2τ of 400 μs for $^{75}$As in Cr-doped semi-insulating GaAs substrate. Applied π-pulses (black arrows) are cancelled out after the add-subtract procedure in phase-cycling technique[5]. Bottom: APCP-echo decay $s_{APCP}$ under homogeneous (left) and inhomogeneous (right) alternating magnetic field for an APCP sequence with 2τ of 200 μs (red), 400 μs (blue) and 600 μs (green) for $^{75}$As in Cr-doped semi-insulating GaAs substrate. Unlike the initial decays which depend on the homogeneity, the slopes in the long-time limit give the same $1/T_2^L$ as shown by the black eye-guides that are parallel in both cases.

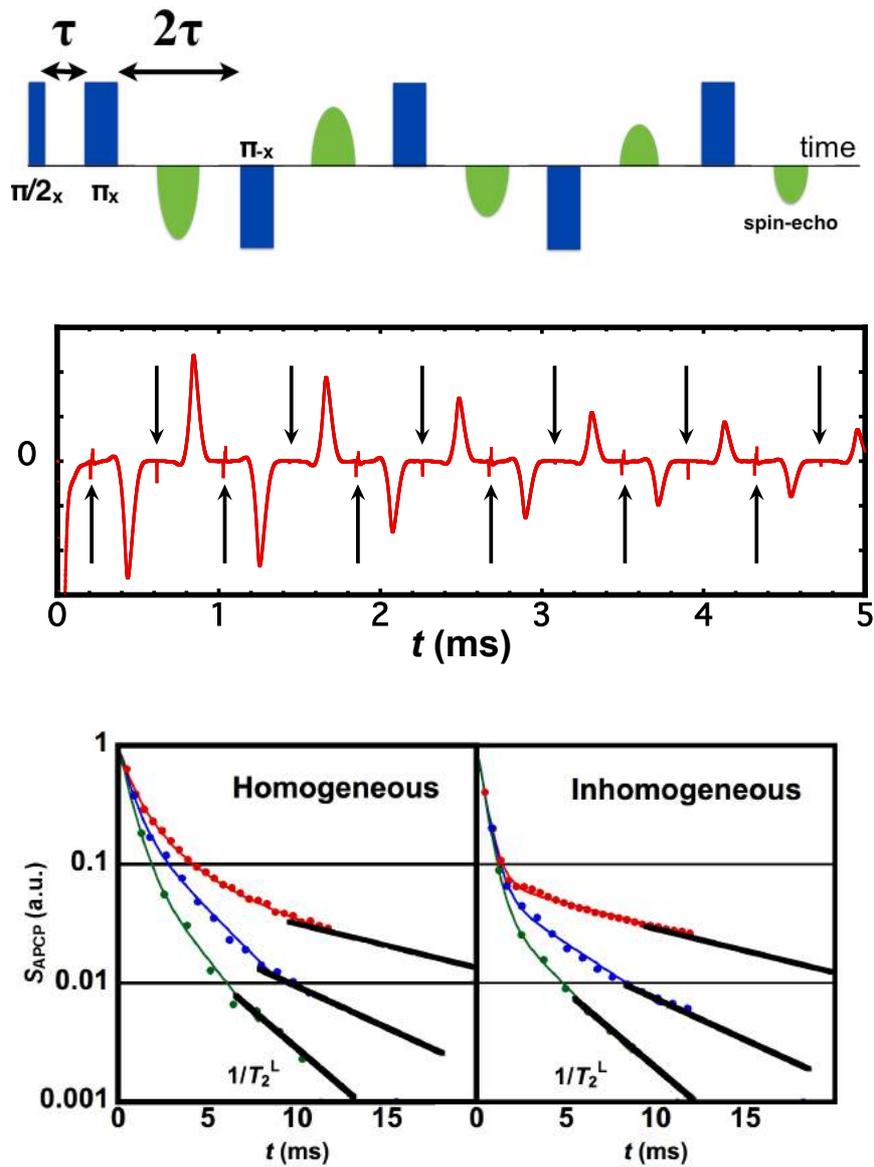

**Figure 2| Multiple $^{75}$As spin-echo decays under APCP pulses along with the Hahn-echo decay.**
(a) un-doped GaAs, (b) Cr-doped semi-insulating GaAs, (c) Si-doped metallic GaAs. In each figure, the data closest to the vertical axis are the Hahn-echo decay, which is obtained by sweeping the pulse interval τ between the first π/2- and the next π-pulse. For shorter and longer pulse intervals, the decays can be fit by two-exponential functions. For intermediate pulse intervals, however, the $T_2^L$s in the two-component fittings do not represent the right slope in the long-time limit[SI-3]. Fitting with three time constants is found to reproduce the decays for all the pulse intervals[SI-4], resulting in the right $T_2^L$s that characterizes the long-time slopes. For the un-doped GaAs, the pulse interval 2 τ are 30 μs, 40 μs, 60 μs, 100 μs, 150 μs, 170 μs, 200 μs, 250 μs, 300 μs, 400 μs, 500 μs and 600 μs. For the Cr-doped GaAs, the pulse interval 2 τ are 30 μs, 40 μs, 60 μs, 100 μs, 150 μs, 170 μs, 200 μs, 240 μs, 300 μs, 400 μs and 600 μs. For the Si-doped GaAs, the pulse interval 2 τ are 30 μs, 40 μs, 60 μs, 100 μs, 150 μs, 170 μs, 200 μs, 250 μs, 300 μs, 400 μs and 600 μs.

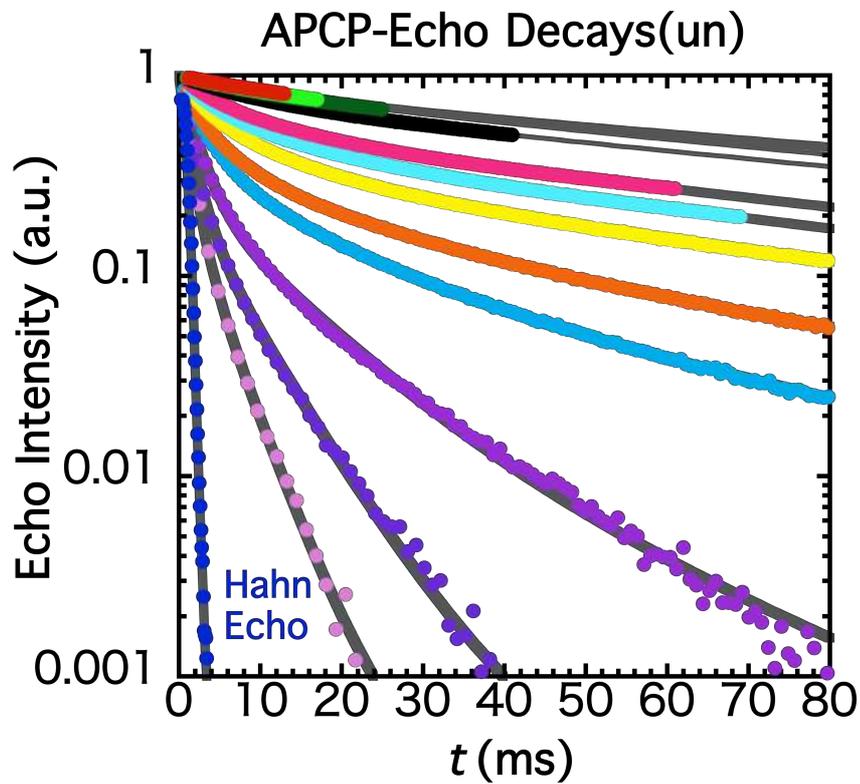

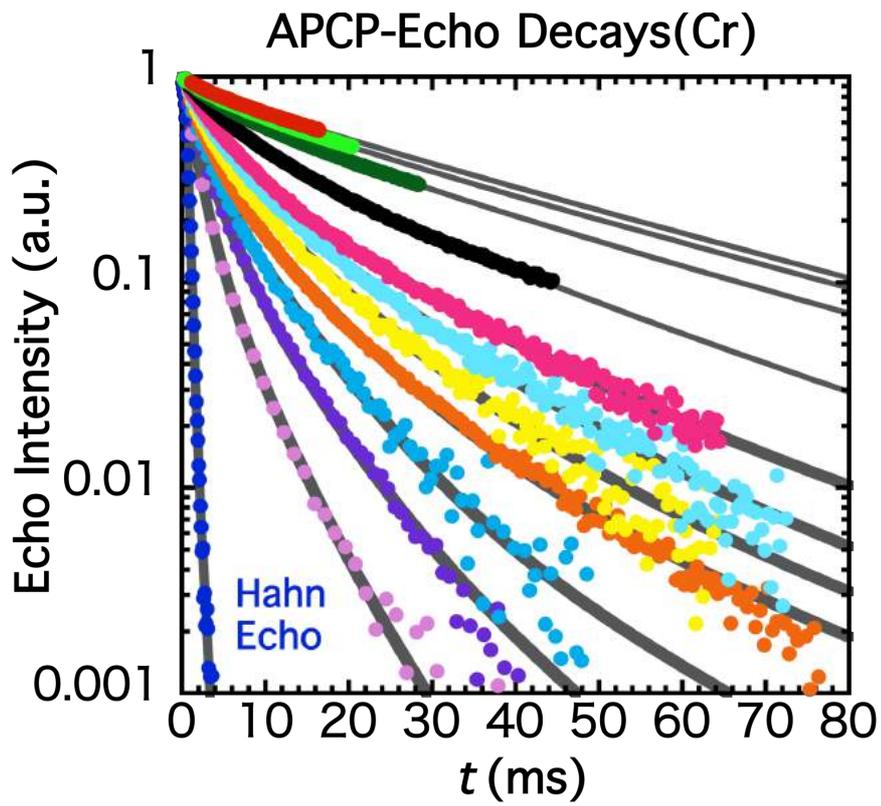

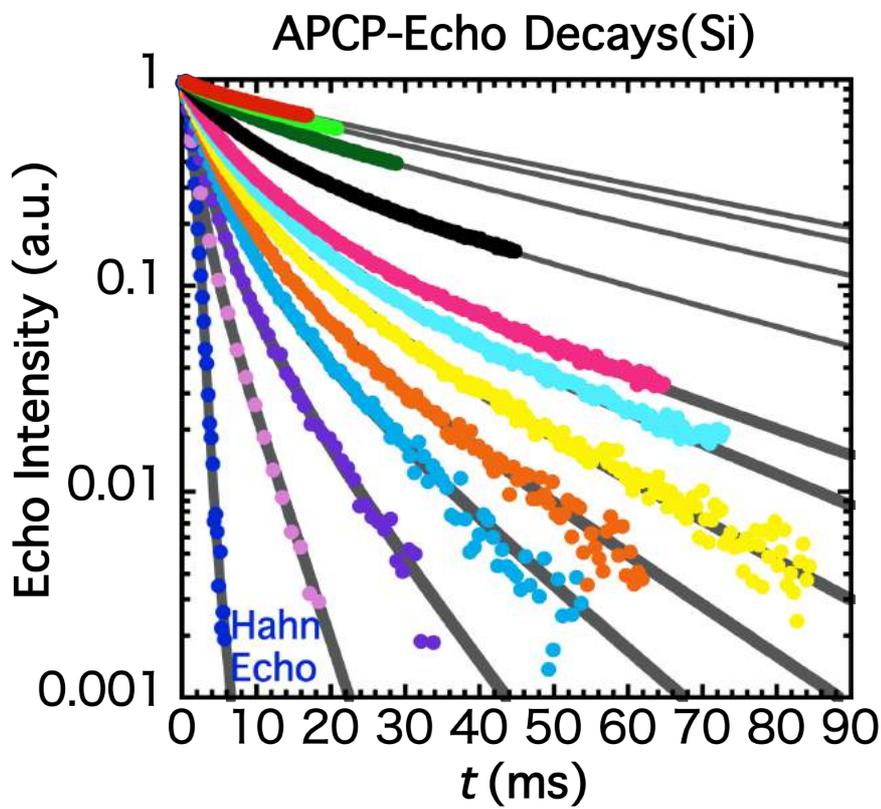

**Figure 3| Spectra of decoherence noise on $^{75}$As nuclear spins in GaAs substrate of (a) un-doped, (b) Cr-doped semi-insulating and (c) Si-doped metallic.** Red circles: $1/T_2^L$ as a function of $1/4\tau$ on $^{75}$As in GaAs substrate at 297 K. Equation (1) and the relation $f=1/4\tau$ validate that the plot is equivalent with the decoherence-noise spectrum $J_\parallel(f)$. The zero-frequency intensities of the spectra are given by the long-time limit of the Hahn-echo decay $T_2$, and hence the $A$ values in Eq. (6). The higher-frequency intensities gives the $B$ values. Given the $A$ and $B$ values, the power $n$ and the characteristic frequency $f_c$ in Eq. (6) are obtained as fitting parameters (solid curves). The frequency dependence is found to be $n=2.00\pm0.01$ for all the substrate with $f_c^{un}=170\pm10$ Hz, $f_c^{Cr}=210\pm10$ Hz and $f_c^{Si}=460\pm30$ Hz. The $1/f^2$ dependence can be seen more clearly by broken curves that fit the blue squares, $B$-term subtracted data. The broken curves exhibit $1/f^2$ dependence over two- or three-orders of magnitude.

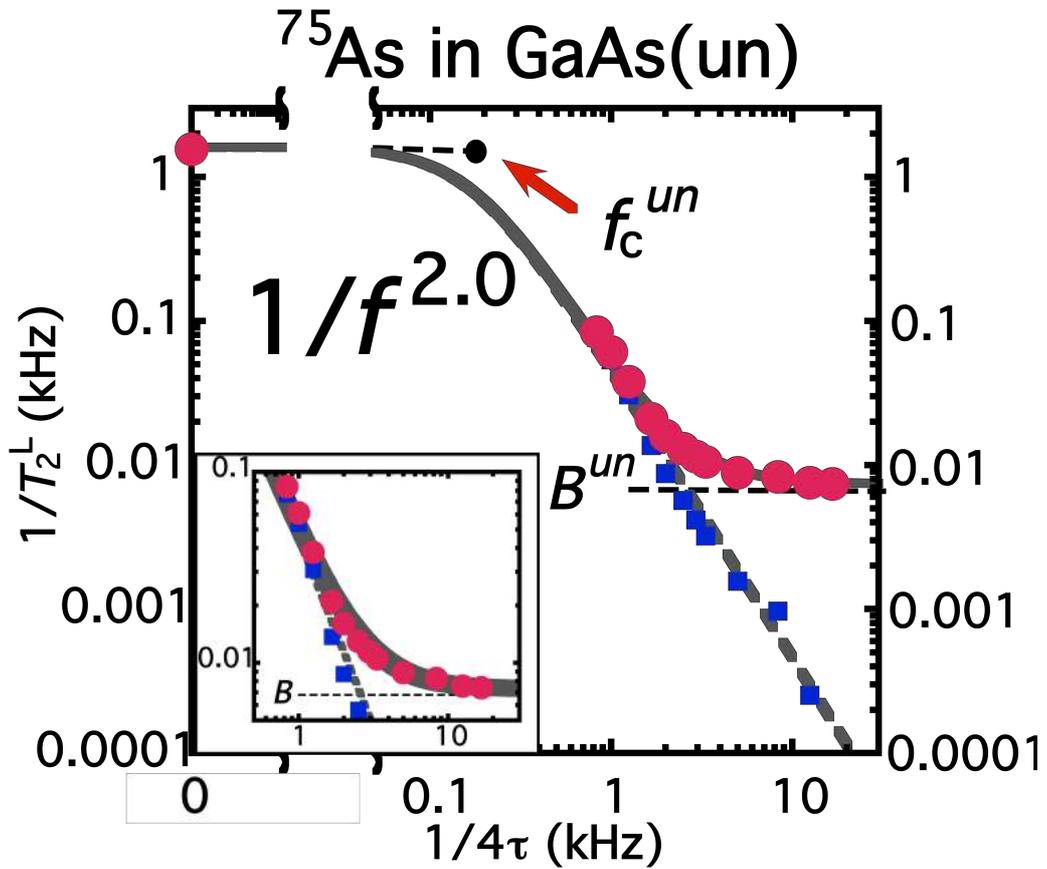

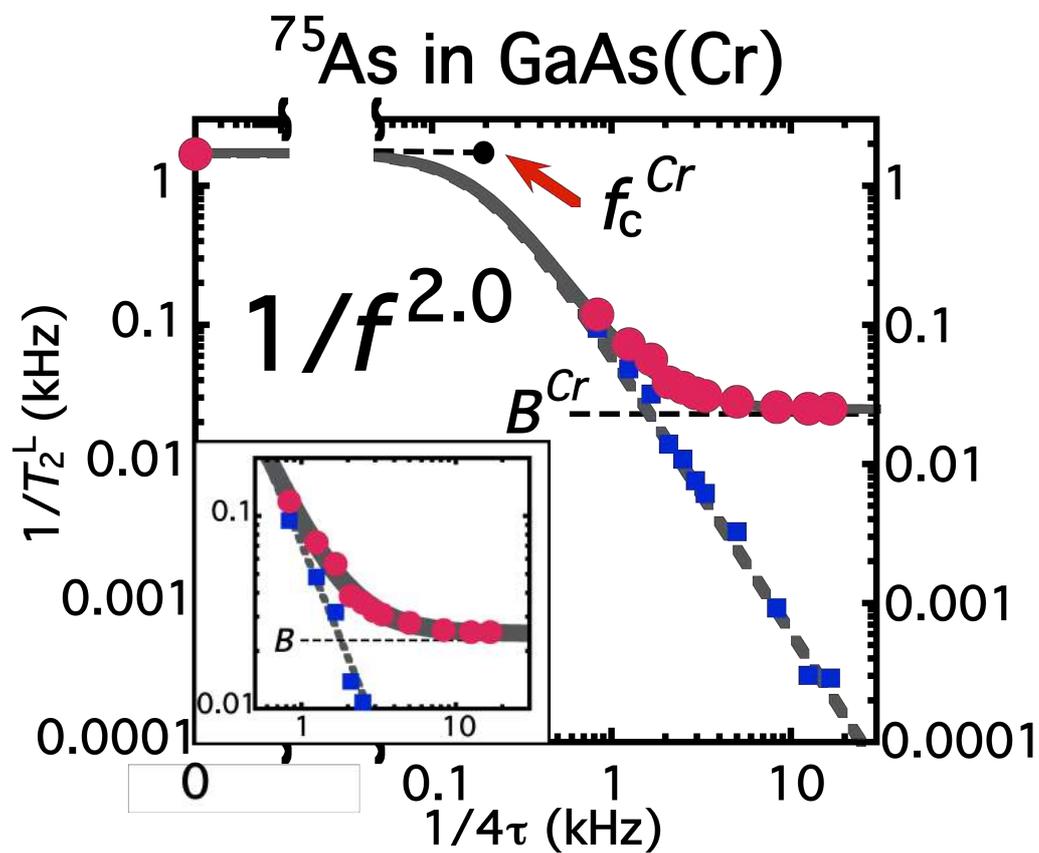

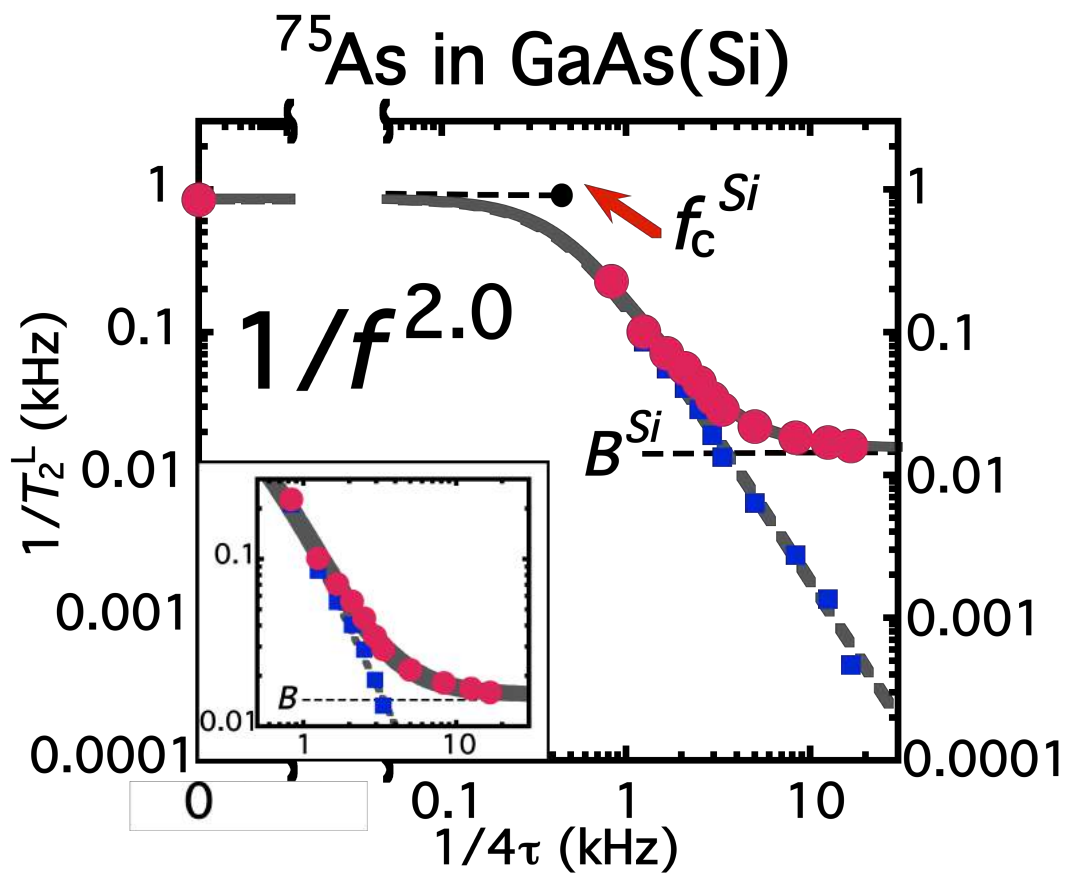

**Figure 4| Noise spectra without the Hahn-echo $T_2$.** Instead of Eq. (6), the spectra are fit by $A/(f/f_c)^n + B$. If the spectral intensities are defined as (a)$T_2^S$, the spectra exhibit the usual "$1/f$" noise such that $n=1.0\pm0.3$. The spectra, where the intensities are defined as (b)$T_2^L$, exhibit $n=2.0\pm0.3$ dependences even in the absence of the Hahn-echo $T_2$. Note that all these spectra stem from the same raw data shown in Fig. 2 as the raw data from which Fig. 3 are deduced.

(a)

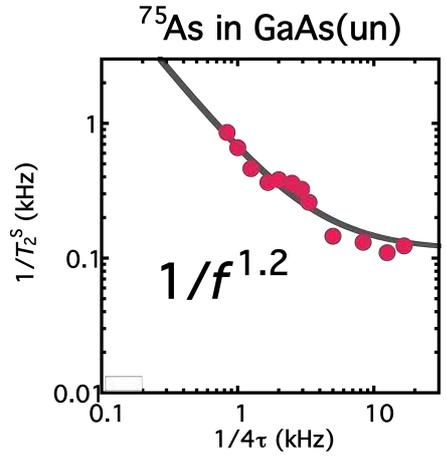

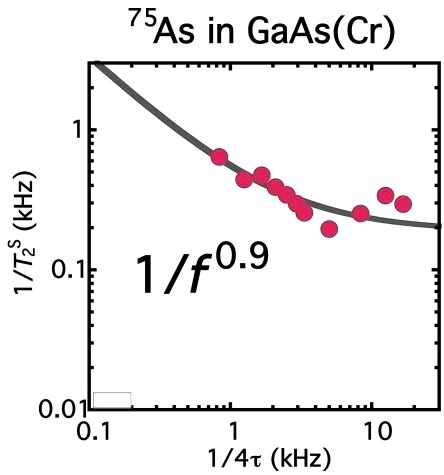

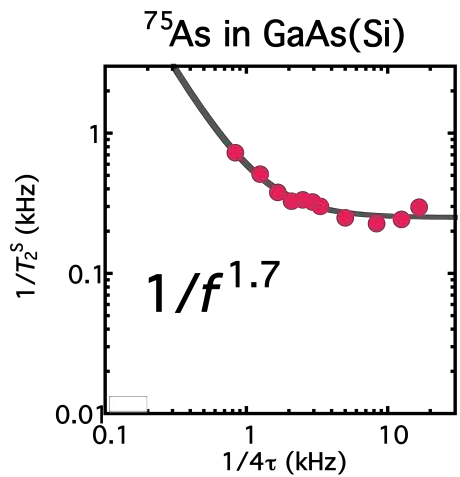

(b)

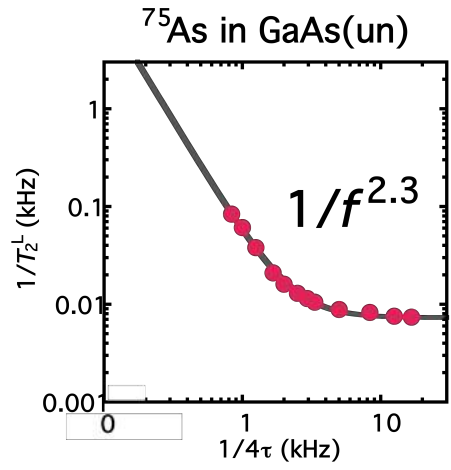

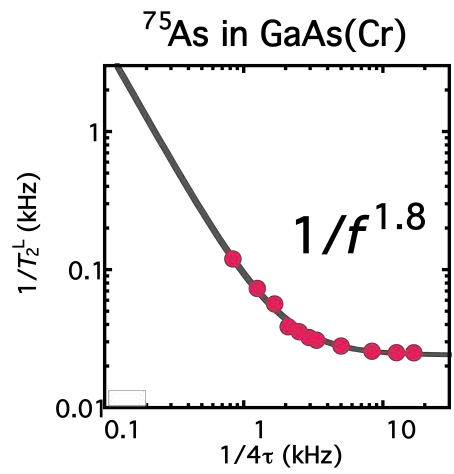

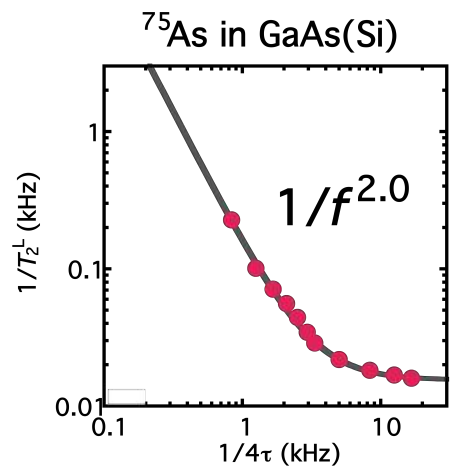

**Figure 5| Noise spectra with the Hahn-echo $T_2$ where the spectral intensities are defined as $T_2^S$.** As in Fig. 4, the spectra are fit by $A/(f/f_c)^n + B$. With the inclusion of the Hahn-echo $T_2$, the spectra are found to exhibit $n=2.0\pm0.3$ dependences even in the definition of the intensities by $T_2^S$. Figs. 4 and 5 validate the definition of the intensities as $T_2^L$ and the inclusion of the Hahn-echo $T_2$, and hence Fig. 3. Note that all these spectra stem from the same raw data shown in Fig. 2 as the raw data from which Fig. 3 are deduced.

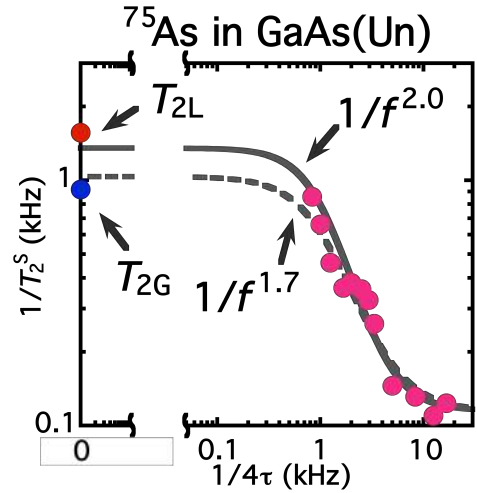

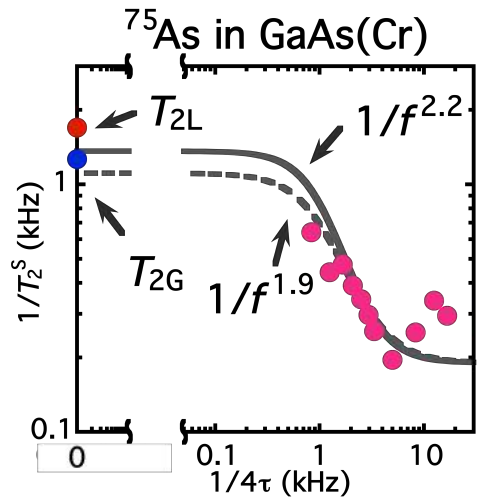

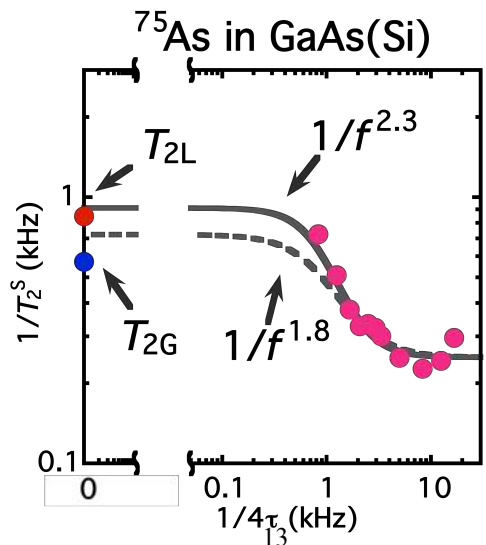

13